\documentclass[runningheads]{llncs}
\usepackage{graphicx}
\usepackage{array}
\usepackage{xcolor}
\usepackage{hyperref}
\usepackage{amsmath}
\begin{document}
\title{Class Activation Map-based Weakly supervised Hemorrhage Segmentation using Resnet-LSTM in Non-Contrast Computed Tomography images}
\titlerunning{Weakly supervised segmentation of hemorrhages}
%
\author{Shreyas H Ramananda\orcidID{0009-0001-3289-5078} \and
Vaanathi Sundaresan\orcidID{0000-0002-9451-4779}\thanks{Corresponding~author; \url{http://cds.iisc.ac.in/faculty/vaanathi/}, Email~id:~vaanathi@iisc.ac.in}}
\authorrunning{S. H. Ramananda et al.}
%
\institute{Department of Computational and Data Sciences, Indian Institute of Science, Bengaluru, Karnataka
560012, India}
\maketitle              
\begin{abstract}
In clinical settings, intracranial hemorrhages (ICH) are routinely diagnosed using non-contrast CT (NCCT) for severity assessment. Accurate automated segmentation of ICH lesions is the initial and essential step, immensely useful for such assessment. However, compared to other structural imaging modalities such as MRI, in NCCT images ICH appears with very low contrast and poor SNR. Over recent years, deep learning (DL)-based methods have shown great potential, however, training them requires a huge amount of manually annotated lesion-level labels, with sufficient diversity to capture the characteristics of ICH. In this work, we propose a novel weakly supervised DL method for ICH segmentation on NCCT scans, using image-level binary classification labels, which are less time-consuming and labor-efficient when compared to the manual labeling of individual ICH lesions. Our method initially determines the approximate location of ICH using class activation maps from a classification network, which is trained to learn dependencies across contiguous slices. We further refine the ICH segmentation using pseudo-ICH masks obtained in an unsupervised manner. The method is flexible and uses a computationally light architecture during testing. On evaluating our method on the validation data of the MICCAI 2022 INSTANCE challenge, our method achieves a Dice value of 0.55, comparable with those of existing weakly supervised method (Dice value of 0.47), despite training on a much smaller training data. 
\end{abstract}
\section{Introduction}
\label{sec:intro}

Acute Intracranial hemorrhage (ICH) is a life-threatening disease, requiring emergency medical attention. Accurate detection and segmentation of ICH could efficiently assist clinicians in assessing their severity. Even though MRI scans show better contrast of ICH, these lesions are routinely diagnosed using non-contrast CT (NCCT) imaging in clinical practice. However, since ICH lesions have poor contrast on CT images with a low signal-to-noise ratio, manual labeling of ICH on CT images is highly time-consuming and labor-intensive. The recent advent of deep learning (DL) models has shown improvement in the detection performance of segmentation tasks on CT images \cite{ref_article1}. However, training DL methods requires large manually labeled datasets, and are affected by variations across hemorrhage sub-types and demographic characteristics (e.g., age). 
\vspace{-0.5em}
\begin{figure}[h!]
    \centering
    \includegraphics[width=\textwidth]{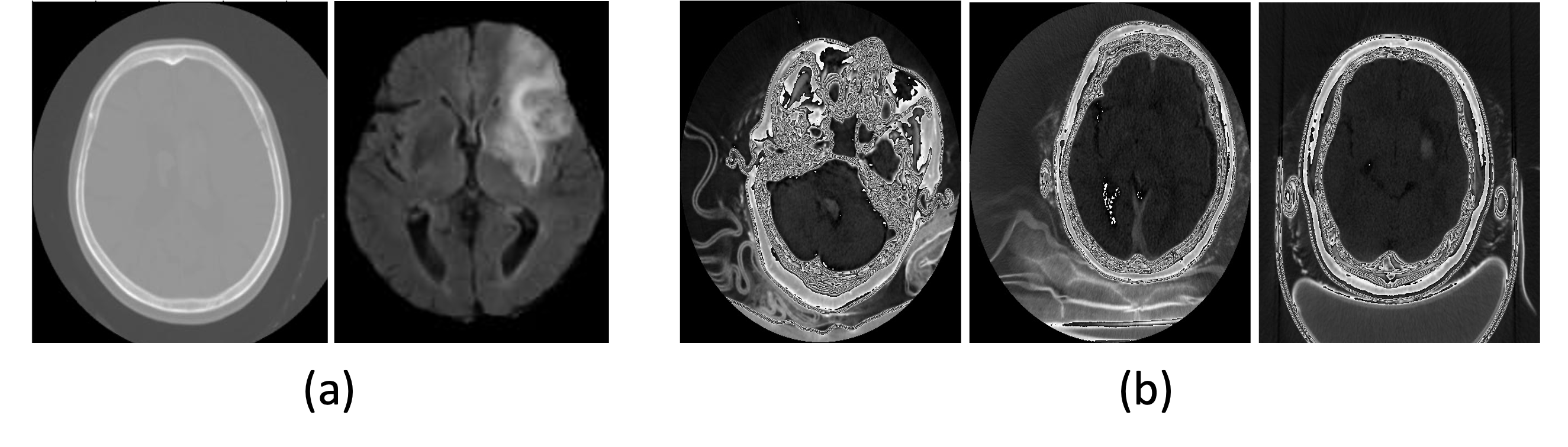}
    \vspace{-2em}
    \caption[]{\textbf{Challenges in CT image segmentation.} (a) Comparison of CT and MRI (b) ICH on RGB-converted CT images }
    \label{fig:intro}
    \vspace{-1em}
\end{figure}
To date, most of the methods for hemorrhage segmentation on CT images are fully supervised. The existing methods for hemorrhage segmentation (at lesion-level) and classification (image-level) have used 3D CNN \cite{ref_article2}\cite{ref_article3} or 2D CNN models \cite{ref_article4}, including encoder-decoder architecture\cite{ref_article5}\cite{ref_article6}, mainly using U-Net \cite{ref_article5}\cite{ref_article6}, and multi-tasking architecture  \cite{ref_article7} for segmentation. 
Moreover, to mitigate the inconsistencies in detection across slices and to reduce classification errors, long short-term memory (LSTM) modules \cite{ref_article8} have been used. A few weakly-supervised methods have been proposed for hemorrhage segmentation on NCCT images to overcome the need for large amount of labelled data. These methods include the use of sliding window (SWIN) transformers \cite{ref_article9} and attention maps \cite{ref_article10} for getting bounding boxes rather than precise boundaries. 
Still, transformers might not be suitable for low-data regimes (where weakly supervised methods would be highly useful), since they require a large amount of training data. They also do not explicitly leverage the spatial information, limiting their ability to capture fine-grained details \cite{ref_article11},\cite{ref_article12}. 
Hence, there is a need for robust weakly supervised methods that would provide accurate detection of hemorrhage on CT images from image-level classification labels. It is also essential for such a method to use spatial contextual information, for efficient quantification and characterization of ICH in real-time clinical applications.

In this work, we aim to accurately localize ICH on NCCT images using a weakly supervised method trained using image-level labels. Our method integrates contextual information across slices in a classification model and uses class activation map \cite{ref_article13} to provide a weak prior for salient locations for hemorrhages. The pseudo-labels obtained in an unsupervised manner from class activation map provides sufficient diversity in their characteristics, thus improving the robustness of the segmentation method. We evaluate our method on a publicly available CT dataset consisting of various sub-types of ICH.
%

\section{Materials \& Methods}
\label{sec:methods}

\begin{figure}[h!]
    \centering
    \includegraphics[width=\textwidth]{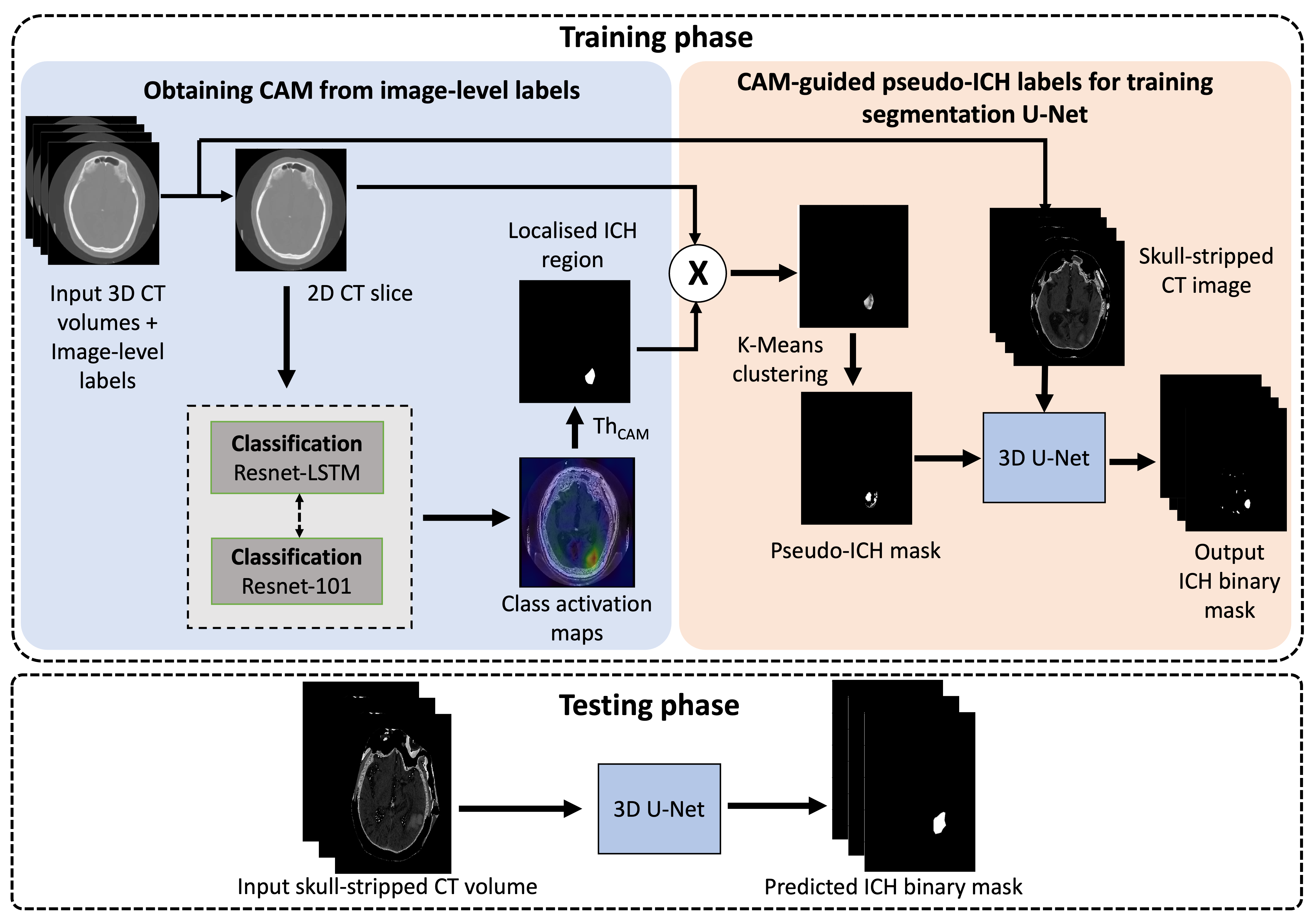}
    \vspace{-1em}
    \caption[]{\textbf{Weakly supervised ICH segmentation from image-level labels. Top:} Training Phase, consisting of 2 steps: obtaining CAM from image-level labels (blue) and CAM-guided pseudo-ICH labels for training 3D U-Net for segmentation (orange). \textbf{Bottom:} Testing phase, where the trained 3D U-Net is used to segment ICH on whole NCCT volumes.}
    \label{fig:method}
\end{figure}

Our main aim is to obtain accurate segmentation for hemorrhages using image-level labels. Towards that aim, our method consists of two steps: (1) obtaining the spatial localization of lesions using the class activation maps (CAM) from initial classification; (2) weakly- supervised refinement of the location and boundary of ICH lesions. The steps in the proposed method are illustrated in Fig.~\ref{fig:method}.\\

\textbf{Getting spatial CAM from image-level labels using ResNet-LSTM:} 
To obtain the location-based information using image-level labels, we train the classification model and estimate the spatially salient regions on the brain that contribute to the image-level decision. We choose the ResNet-101 \cite{ref_article14} on the input 2D slices since it has been successfully used for classification tasks in medical imaging \cite{ref_article15}. The input slices are converted from grayscale to RGB colorspace (as shown in Fig.~\ref{fig:intro}) to be provided as input to 2D ResNet. The main reason for using 2D ResNet is to provide negative samples for training since all subjects had hemorrhages (and hence the lack of negative samples in 3D volumes). Since the ResNet model is trained on the slices individually, to ensure the continuity across slices, we integrate an LSTM (after convergence of ResNet) and continue training again until convergence. The LSTM establishes short-term dependencies across slices ensuring continuity. Hence, to obtain the 3D CAM (with continuity across slices), we integrate the LSTM module with the ResNet by removing the fully connected (FC) layers. After training, we remove the LSTM module, since the weights of the ResNet backbone are updated to learn the short-term dependencies across slices. We later add the FC layers and obtained contextually meaningful 3D CAM, which provides localized feature saliency, highlighting the potential hemorrhage regions. We use binary cross-entropy loss function for true target value $y$ and predicted value $\hat{y}$ to train the classification model as follows:
\begin{equation}
\text{BCE}(y, \hat{y}) = -\left(y \log(\hat{y}) + (1 - y) \log(1 - \hat{y})\right)
\label{eqn:bce}
\end{equation}

\textbf{Obtaining pseudolabel masks for segmentation:} Since the CAM provide a weak prior of hemorrhage location, we threshold the CAM at the threshold Th$_{CAM}$ to obtain the region containing hemorrhage. The region within the threshold CAM region is clustered using an unsupervised K-Means algorithm. To select a hemorrhage cluster from others, we use differential prediction as follows: (i) From the K-means clustering output, we subtract the region corresponding to each cluster from the original image, (ii) we feed the subtracted image to the trained ResNet, (iii) among all the clusters, we take the cluster corresponding to the lowest value of logits as ICH cluster (since subtracting the ICH-cluster would lead to loss of information and hence is expected to provide the lowest logits value). We consider this cluster mask as a pseudo-lesion mask for weakly supervised hemorrhage segmentation in the next step. \\

\textbf{CAM-guided Weakly supervised segmentation using pseudo-ICH labels: }
The main aim for this step is to improve the hemorrhage localization and improve its segmentation. Hence, we perform  {contrast enhancement} and remove the skull areas to avoid spurious bright regions that could lead to erroneous detection. For the skull removal, we smooth the image using bilateral filtering \cite{ref_article16} and threshold the image at 95$^{th}$ percentile of intensity values. We discard the skull and anything outside the convex hull of the skull (example shown in Fig.~\ref{fig:method}). We train the 3D U-Net \cite{ref_article17} model in a supervised manner by using the pseudo-label masks obtained in the above step. We use a combination of BCE and Dice loss function for training the U-Net model for target mask $y$ and predicted mask $\hat{y}$. \
\begin{equation}
L_{comb}(y, \hat{y}) = (1 - \frac{2 \sum (y \odot \hat{y}) + \epsilon}{\sum y + \sum \hat{y} + \epsilon}) + (-(y \log(\hat{y}) + (1 - y) \log(1 - \hat{y})))
\label{eqn:combination_loss}
\end{equation}

\textbf{Testing phase:} During the testing phase, we perform skull extraction and histogram equalization and apply the trained 3D U-Net to get the final ICH segmentation mask on 3D NCCT volumes.

\subsection{Implementation details}
\label{ssec:imp_details}

For training ResNet-101 and ResNet-LSTM, we use the SGD optimizer with momentum=0.9, with a batch size of 32 and a learning rate of 0.0001 (0.001 from ResNet-LSTM) for 100 epochs with a patience value of 25 epochs for early stopping (converged at $\sim$45 epochs). 
We use a threshold value of 0.7 on CAM, and for K-Means clustering, the value of K is 4. We train 3D U-Net using Adam optimizer \cite{ref_article18} with a batch size of 4 and learning rate of 0.001 with the same number of epochs and patience value as mentioned above. The above hyperparameters are chosen empirically.

\section{Experiments}
\label{sec:exps}

\subsection{Dataset Details}
\label{ssec:data} 
We use 100 3D CT scans released as training data for the MICCAI 2022 ICH segmentation challenge (\url{https://instance.grand-challenge.org/}, INSTANCE) \cite{ref_article19} on NCCT images, including 5 different types of hemorrhages: intraparenchymal hemorrhage (IPH), intraventricular hemorrhage (IPH), subarachnoid hemorrhage (SAH), subdural hemorrhage (SDH), and epidural hemorrhage (EDH). Manual lesion labels are available for the scans. However, since we aim to develop a weakly supervised method that could provide segmentation results with image-level labels, we extract image-level classification labels from the manual lesion labels (present/positive sample if any sub-type of the lesion are labeled in a slice, absent/negative otherwise). The volumes are anisotropic, with dimensions ranging from 512 $\times$ 512 $\times$ 20 to 512 $\times$ 512 $\times$ 70, and a voxel resolution of 0.42mm $\times$ 0.42mm $\times$ 5mm.\\

\textbf{Experiments: }
We submitted our method to the INSTANCE challenge dataset to be evaluated in-house by the organizers on the unseen validation data. We also compare with a fully supervised baseline on the same dataset, and another weakly supervised method \cite{ref_article9} on a different CT dataset. Additionally, on the INSTANCE training data that is publicly available, we perform 3-fold cross-validation with a training-validation-test split of 70-10-20 subjects (2240 and 320 slices for ResNet-LSTM training and validation respectively). We also perform an ablation study to investigate the effect of various components: (1) using the CAM from ResNet alone as pseudo-ICH labels for 3D U-Net training (ResNet + U-Net), (2) using the CAM  from ResNet-LSTM as pseudo-ICH labels for 3D U-Net (ResNet-LSTM + U-Net), (3) Using ResNet CAM and K-means clustering to get pseudo-ICH labels for 3D U-Net training (ResNet + K-means + U-Net), (4) Using ResNet-LSTM CAM and K-means clustering to get pseudo-ICH labels for 3D U-Net training (ResNet-LSTM + K-means + U-Net). We also determine the statistical significance of results using two-tailed, paired t-tests.

\hspace{1em} \textbf{Performance evaluation metrics:} 
We use (1) Dice overlap measure (Dice), (2) relative volume difference (RVD):$\frac{|V_{\text{pred}} - V_{\text{gt}}|}{V_{\text{gt}}}$, where $V_{\text{pred}}$ and $V_{\text{gt}}$ are predicted output and ground truth volumes respectively, (3) voxel-wise true positive rate (TPR). Also, Hausdorff distance (HD) and surface Dice (SD) values were additionally determined by the organizers on submitting our method to INSTANCE challenge validation data.
\\

\section{Results and discussion}
\label{sec:results}
\textbf{Validation results and comparison with existing work: }Evaluation results of our method on the unseen INSTANCE validation data by the challenge organizers are reported in table~\ref{table:val_comp}, along with the comparison with existing methods. On comparing our method with another weakly supervised method using SWIN transformer \cite{ref_article9} for binary classification, our Dice values (0.55) on the INSTANCE validation data compares favourably with the Dice value of 0.47 obtained by \cite{ref_article9} on PhysioNet dataset \cite{ref_artic20}, despite using only 2240 training slices (from 70 subjects) for training (20,000 samples used by \cite{ref_article9}). This could be due to the fact that we also take 3D contextual information into consideration using LSTM. 
A fully supervised U-Net \cite{ref_article20}, used as a baseline in INSTANCE challenge \cite{ref_article19} provided a Dice value of 0.64. Comparing this with our results, our method shows the potential to be able to provide comparable results with a larger and more diverse training data.

\begin{table}
\centering
\scriptsize
\caption{Performance on unseen validation data from the challenge and comparison with existing literature.}
\label{tab1_transposed}
\begin{tabular}{|l|l|l|l|l|l|}
\hline
\textbf{Methods} & \textbf{Dataset} &\textbf{DICE} & \textbf{RVD} & \textbf{HD} & \textbf{SD} \\
\hline
\textbf{Our Method} & \parbox[c][][t]{1.5cm}{\vspace{3pt} \raggedright INSTANCE validation (unseen) \vspace{2pt}}& $\mathbf{0.55 \pm 0.3}$ & $\mathbf{0.79 \pm 1.22}$ & $\mathbf{42.75 \pm 32.54}$ & $\mathbf{0.28 \pm 0.17}$ \\
\hline
\parbox[c][][t]{2.0cm}{\raggedright \textbf{Fully sup U-Net(baseline) [14]}\\[3pt]} & \parbox[c][][t]{1.5cm}{\vspace{3pt} \raggedright INSTANCE validation (unseen)\\[3pt]} & $0.64 \pm 0.27$ & $0.46 \pm 0.2$ & $277.63 \pm 163$ & $0.51 \pm 1.14$ \\
\hline
\parbox[c][][t]{2.0cm}{\vspace{3pt} \raggedright \textbf{Weakly supervised[8] } \vspace{2pt}} & PhysioNet & $0.47 \pm 0.26$ & $-  $ & $- $& $- $  \\
\hline
\end{tabular}
\label{table:val_comp}
\end{table}

\begin{figure}[h!]
    \centering
    \includegraphics[width=\textwidth]{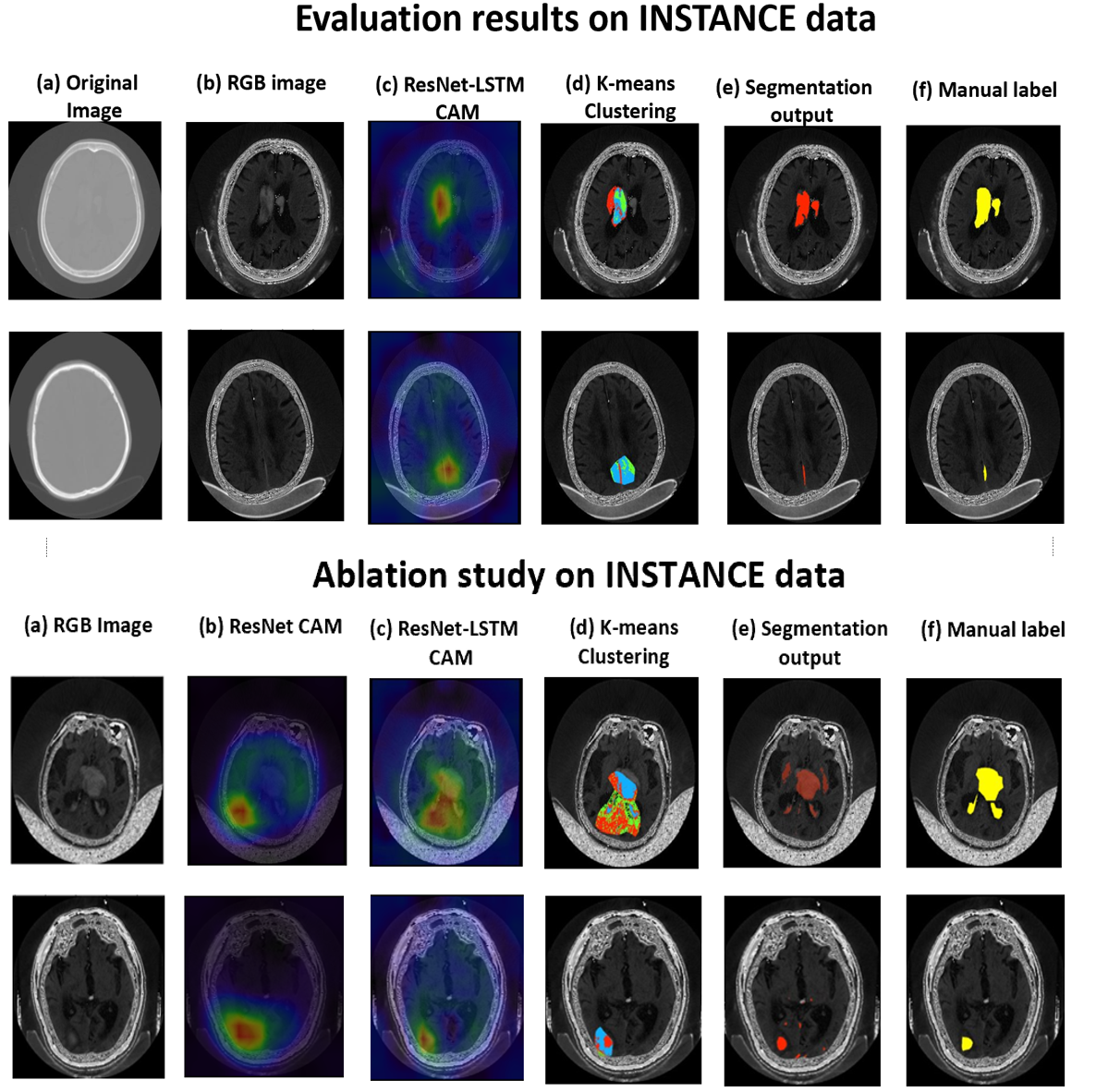}
    \vspace{-1em}
    \caption[]{\footnotesize{\textbf{Evaluation on INSTANCE data and ablation study. Top panel:} Results shown for NCCT 2D slices (b), along with CAM maps from ResNet-LSTM classification model (c), K-means clusters from CAM (d) and segmentation output (e). \textbf{Bottom panel:} CAM maps from ResNet alone (b) shown along with CAM from ResNet-LSTM (c); K-means clustering output (d) and segmentation output (e) are shown corresponding to CAM map from (c). In both top and bottom panels, manual segmentations are shown in (f).}}
    \label{fig:results_visual}
\end{figure}

\textbf{Cross-validation on publicly available INSTANCE data: }The results of 3-fold cross validation are reported in table~\ref{table:cv_abl} and the visual results are shown in Fig~\ref{fig:results_visual}.
On performing the cross-validation, our method achieve a mean Dice of 0.47 and a mean RVD of 1.05 at the threshold of 0.5. As shown in Fig.~\ref{fig:results_visual} (top panel), CAM  provides a good estimate of ICH location, while the boundaries are further improved by the 3D U-Net segmentation model. 

\begin{table}
\centering
\caption{Cross Validation Performance and Ablation study on the INSTANCE data (*/** indicate values significantly above those in the previous row using a paired t-test; *p-value$<$0.05, **p-value$<$0.001)}
\label{tab1_transposed}
\begin{tabular}{|l|l|l|l|l|}
\hline
\textbf{Metric} & \textbf{DICE} & \textbf{RVD} & \textbf{TPR}  \\
\hline
Cross-validation & $0.47 \pm 0.3$ & $1.05 \pm 1.22$ & $0.48$  \\
\hline
\multicolumn{4}{|c|}{\textbf{Ablation Study}} \\
\hline
ResNet + U-Net &$ 0.14 \pm  0.17$ & $3.15 \pm 3.69$ & $0.11$\\
ResNet-LSTM + U-Net & $0.31 \pm 0.23^{**}$ & $1.54 \pm 1.37^*$ & $0.32^*$ \\
\hline
ResNet + K-means + U-Net & $0.18 \pm 0.2$ & $0.82 \pm0.54 $ & $0.24$  \\
ResNet-LSTM + K-means + U-Net & $0.57 \pm0.26^{**} $ & $0.37 \pm 0.32^{**}$ & 0.48$^*$ \\
\hline
\end{tabular}
\label{table:cv_abl}
\end{table}


\textbf{Ablation study results: }From table~\ref{table:cv_abl} and the bottom panel of Fig.~\ref{fig:results_visual}, out of all the settings, ResNet + U-Net has the worst performance with a mean Dice of 0.14. This is also evident from the spatially inaccurate CAM that we obtain from the ResNet model (c). Integrating LSTM for fine-tuning the ResNet classification model provides significant improvement in the spatial localization accuracy of CAM  (in terms of size, focus, and location as shown in d) and hence the segmentation output (e) since it takes into consideration the context across slices. Also, from the paired t-test results (table~\ref{table:cv_abl}), the performance of our method significantly improves with the integration of LSTM within the ResNet model. This is in line with the result of the INSTANCE challenge, where most of the top-performing methods consider both 2D and 3D information. \\
Out of all methods, the ResNet-LSTM + K-means + U-Net model performs the best with a mean Dice value of 0.57, indicating the use of clustered rough pseudo-labels in leveraging the difference between ICH and background regions (also evident from the improvement in segmentation results in Fig.~\ref{fig:results_visual}e). The main sources of false positives are the high-intensity regions near skull areas and confounding structures similar to lesions. However, our contrast enhancement during preprocessing (shown in Fig.~\ref{fig:method}) improves the contrast of CT images thus efficiently aiding the models in learning discriminative features for ICH concerning the background.

Currently, our method provides lower performance while segmenting thinner lesions closer to the skull (e.g., SAH and EDH), thus affecting our overall performance. One of the future directions of our work would be to improve the segmentation accuracy and make the method generalizable across various subtypes of ICH. 

\section{Conclusion}
Our proposed weakly supervised method is trained for ICH segmentation using image-level labels. Our method is highly flexible and computationally light during prediction. Our method takes into consideration both 2D and 3D features contributing to the classification decision, by obtaining CAM  from ResNet fine-tuned using LSTM. Our method achieves a Dice value of 0.55 on the validation data of MICCAI 2022 INSTANCE challenge and performs on par with the existing weakly supervised method for ICH segmentation.  

\section*{Acknowledgements}
The authors acknowledge the Start-up grant from Indian Institute of Science (IISc), India.

\label{sec:ack}

%
%
%

\begin{thebibliography}{8}
    \bibitem{ref_article1}
    Baig, R., et al. “Deep Learning Approaches Towards Skin Lesion Segmentation and Classification from Dermoscopic Images - A Review.” Current medical imaging vol. 16,5 (2020): 513-533. doi:10.2174/1573405615666190129120449
    \bibitem{ref_article2}
    Patel, A., et al. “Intracerebral Haemorrhage Segmentation in Non-Contrast CT.” Scientific reports vol. 9,1 17858. 28 Nov. 2019, doi:10.1038/s41598-019-54491-6
    \bibitem{ref_article3}
    Singh, S.P., et al. “Shallow 3D CNN for Detecting Acute Brain Hemorrhage From Medical Imaging Sensors.” IEEE Sensors Journal 21 (2021): 14290-14299.
    \bibitem{ref_article4}
    Wang, X., et al. “A deep learning algorithm for automatic detection and classification of acute intracranial hemorrhages in head CT scans.” NeuroImage. Clinical vol. 32 (2021): 102785. doi:10.1016/j.nicl.2021.102785
    \bibitem{ref_article5}
    Lu, L., et al., "Deep Learning for Hemorrhagic Lesion Detection and Segmentation on Brain CT Images." in IEEE Journal of Biomedical and Health Informatics, vol. 25, no. 5, pp. 1646-1659, May 2021, doi: 10.1109/JBHI.2020.3028243.
    \bibitem{ref_article6}
    Schraut, JX., Liu, L.,Gong, J., et al. "A multi-output network with U-net enhanced class activation map and robust classification performance for medical imaging analysis." Discov Artif Intell. 2023;3(1):1. doi: 10.1007/s44163-022-00045-1.
    \bibitem{ref_article7}
    Guo, D., Wei, H., Zhao, P., et al. "Simultaneous Classification and Segmentation of Intracranial Hemorrhage Using a Fully Convolutional Neural Network." In: Proceedings of the 2020 IEEE International Symposium on Biomedical Imaging (ISBI). 2020;118-121. doi: 10.1109/ISBI45749.2020.9098596.
    \bibitem{ref_article8}
    Nguyen, N., et al. "A CNN-LSTM architecture for detection of intracranial hemorrhage on CT scans." medRxiv (2020): 2020-04.
    \bibitem{ref_article9} 
    Rasoulian, A., et al. "Weakly Supervised Intracranial Hemorrhage Segmentation using Head-Wise Gradient-Infused Self-Attention Maps from a Swin Transformer in Categorical Learning." arXiv preprint arXiv:2304.04902 (2023).
    \bibitem{ref_article10}
    Nemcek, J., Vicar, T., and Jacubikek, R. "Weakly supervised deep learning-based intracranial hemorrhage localization." arXiv preprint arXiv:2105.00781 (2021).
    \bibitem{ref_article11} 
    He, K., et al. "Transformers in medical image analysis." Intelligent Medicine 3.1 (2023): 59-78.
    \bibitem{ref_article12} 
    Fahad, S., et al. "Transformers in medical imaging: A survey." Medical Image Analysis (2023): 102802.
    \bibitem{ref_article13}    
    Zhou, B., et al. "Learning deep features for discriminative localization." Proceedings of the IEEE conference on computer vision and pattern recognition. 2016.
    \bibitem{ref_article14}
    He, K., et al. "Deep residual learning for image recognition." Proceedings of the IEEE conference on computer vision and pattern recognition. 2016.
    \bibitem{ref_article15}
    Ghosal, P., Nandanawar, L., et al. "Brain Tumor Classification Using ResNet-101 Based Squeeze and Excitation Deep Neural Network." 2019 Second International Conference on Advanced Computational and Communication Paradigms (ICACCP), Gangtok, India, 2019, pp. 1-6, doi: 10.1109/ICACCP.2019.8882973.
    \bibitem{ref_article16}
    Tomasi, C., Manduchi, R. "Bilateral filtering for gray and color images." Sixth International Conference on Computer Vision (IEEE Cat. No.98CH36271), Bombay, India, 1998, pp. 839-846, doi: 10.1109/ICCV.1998.710815.
    \bibitem{ref_article17}    
    Ronneberger, O., Fischer, P., et al. "U-net: Convolutional networks for biomedical image segmentation." Medical Image Computing and Computer-Assisted Intervention–MICCAI 2015: 18th International Conference, Munich, Germany, October 5-9, 2015, Proceedings, Part III 18. Springer International Publishing, 2015.
    \bibitem{ref_article18}    
    Kingma, D.P., Ba, J., "Adam: A method for stochastic optimization." arXiv preprint arXiv:1412.6980 (2014).
    \bibitem{ref_article19}
    Li, X., et al. "The state-of-the-art 3D anisotropic intracranial hemorrhage segmentation on non-contrast head CT: The INSTANCE challenge." arXiv preprint arXiv:2301.03281 (2023).
    \bibitem{ref_artic20}
    Hssayeni, M.D., et al. "Intracranial hemorrhage segmentation using a deep convolutional model." Data 5.1 (2020): 14.
    \bibitem{ref_article20}
    X, Li., et al. “Hematoma Expansion Context Guided Intracranial Hemorrhage Segmentation and Uncertainty Estimation.” IEEE journal of biomedical and health informatics vol. 26,3 (2022): 1140-1151. doi:10.1109/JBHI.2021.3103850

\end{thebibliography}
%

\end{document}